\documentclass[cits,a4paper]{PoS}
\usepackage{iftex}
\ifXeTeX
  \usepackage{fontspec}
\fi
\usepackage{subcaption}
\usepackage{amsmath}

\ifXeTeX
\fi

\newcommand{\figref}[1]{figure~\ref{#1}}
\newcommand{\Figref}[1]{Figure~\ref{#1}}
\renewcommand{\eqref}[1]{eq.~\ref{#1}}

\newcommand{\la}{\langle}
\newcommand{\ra}{\rangle}
\DeclareMathOperator{\trace}{Tr}

\title{Strange quark momentum fraction from overlap fermion}
\ShortTitle{Strange quark momentum fraction from overlap fermion}
\author{\speaker{Mingyang Sun} \\
  Department of Physics and Astronomy, University of Kentucky,
  Lexington, KY 40506 \\
  E-mail: \email{mingyang.sun@uky.edu}}
\author{Yi-Bo Yang \\
  Department of Physics and Astronomy, University of Kentucky,
  Lexington, KY 40506 \\
  E-mail: \email{ybyang@pa.uky.edu}}
\author{Keh-Fei Liu \\
  Department of Physics and Astronomy, University of Kentucky,
  Lexington, KY 40506 \\
  E-mail: \email{liu@pa.uky.edu}}
\author{Ming Gong \\
  Institute of High Energy Physics, Chinese Academy of Sciences,
  Beijing 100049, China \\
  E-mail: \email{gongming@ihep.ac.cn}

  \begin{center}
    \large{
      \vspace*{0.4cm}
      \includegraphics[scale=0.20]{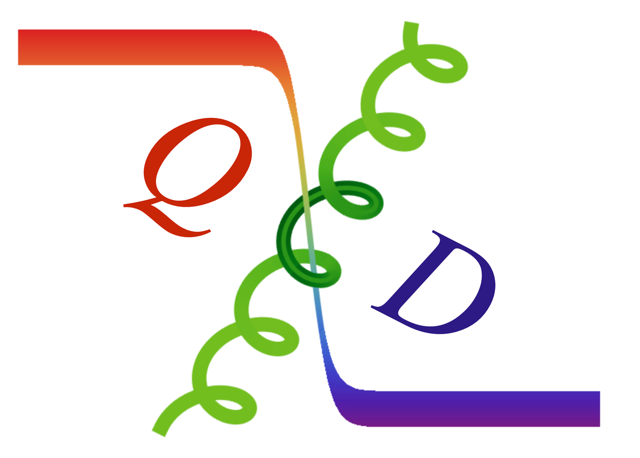}\\
      \vspace*{0.4cm}
      ($\chi$QCD Collaboration)
    }
  \end{center}
}

\abstract{We present a calculation of $\langle x \rangle_s$ for the
  strange quark in the nucleon. We also report the ratio of the strange $\langle x \rangle$ to that of
  $u/d$ in the disconnected insertion which will be useful in constraining the
  global fit of parton distribution functions at small $x$.  We adopt overlap fermion action on
  $2  + 1$ flavor domain-wall fermion configurations on the
  $24^3 \times 64$ lattice with a light sea quark mass which corresponds to $m_{\pi}=330$ MeV.
  Smeared grid $Z_3$ sources are deployed to calculate the nucleon propagator with low-mode
  substitution.  Even-odd grid sources and time-dilution technique
  with stochastic noises are used to calculate the high mode
  contribution to the quark loop. Low mode averaging (LMA) for the quark loop is applied to
  reduce the statistical error of the disconnected insertion
  calculation. We find the ratio  $\langle x \rangle_s/\langle x \rangle_{u/d}^{\mathrm{DI}}= 0.78(3)$
   in this study.}

 \FullConference{The 32nd International Symposium on Lattice Field
   Theory \\
   June 22 -- June 29, 2014 \\
   Columbia University \\
   New York, USA}

\begin{document}

\section{Introduction}

The understanding of the structure of nucleon has been one of the
central issues in hadron physics. For instance, the parton
distribution functions (PDFs) have been studied extensively, and the
observation of scaling violation in PDFs provides the cradle for the
establishment of the fundamental theory, QCD. Yet, there exist many
unresolved questions for the structure of the nucleon.  Ever since the
EMC experiment showed that the proton spin carried by quarks is
small\cite{ashman_measurement_1988}, large effort has been made in both
the experiment and the theoretical frontiers to identify all the
contributions to the nucleon spin.  Calculating the momentum fraction
$\la x \ra$ is an integral part of the study of this subject.

However, it has been found that calculating the disconnected insertion
contribution to this quantity, which is necessary for the strange
momentum fraction, is extremely difficult.  Most lattice calculations
are done for the connected insertion contribution.  So far, only one
quenched calculation exists that includes all contributions to the
nucleon spin\cite{m._deka_moments_2009}.  In this paper, we calculate
the strange momentum fraction with the overlap fermion on $2+1$ flavor
domain wall fermion configurations with the help of an array of
lattice techniques.  We also calculate the momentum fraction in the
u/d channel for the disconnected insertion (DI), and take the ratio
between strange and u/d channel in DI.  In the end, we compare our
result with previous lattice calculations, as well as current global
analyses of parton distribution at small $x$.

\section{Formalism and simulation parameters}

The momentum fraction carried by quarks can be obtained by calculating
the following quark operator of the QCD energy-momentum tensor in the forward matrix
element of the nucleon at finite momentum.
\begin{equation}
  T_{4i} = - \frac{i}{2} \left(\bar{q} \gamma_{4}
    \stackrel{\leftrightarrow}{D}_{i} q + \bar{q} \gamma_{i}
    \stackrel{\leftrightarrow}{D}_{4} q\right)
\end{equation}
in which
\begin{displaymath}
  \gamma_{i} \stackrel{\leftrightarrow}{D}_{j} = \frac12 \left( \gamma_{i}
    \overrightarrow{D}_{j} - \gamma_{i} \overleftarrow{D}_{j} \right).
\end{displaymath}

\begin{figure}
  \centering
  \includegraphics[width=0.3\textwidth]{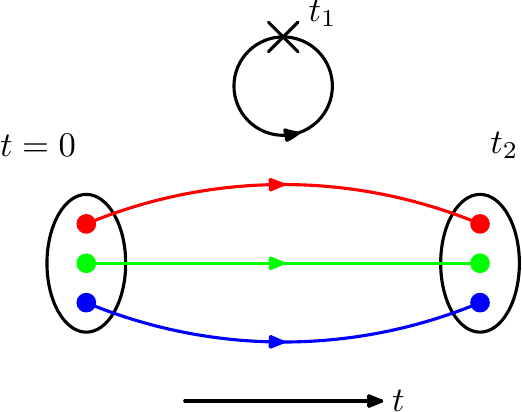}
  \caption{The disconnected insertion}
  \label{fig:di}
\end{figure}

With this operator, we calculate the disconnected three-point function

\begin{equation}
  C_3(t_1, t_2, \vec{p}) = \la P_N (t_2, \vec{p}) L_{4i}(t_1) \ra -
    C_2^N (t_2, \vec{p}) \la L_{4i}(t_1) \ra, \quad t_1 < t_2
\end{equation}
in which the angular bracket denotes ensemble average, $P_N$ is the nucleon propagator, and $L$
denotes the energy-momentum tensor contracted in a quark loop.  The
momentum-projected nucleon two-point function $C_2^N$ is
\begin{equation}
  C_2^N(t_2, \vec{p}) = \la P_N(t_2, \vec{p})\ra = \sum_{\vec{x}} e^{-i \vec{p} \cdot \vec{x}} \la 0 |
  T[\chi(\vec{x}, t_2) \bar{\chi}(0, 0)]|0 \ra.
\end{equation}
where $\chi(\vec{x}, t)$ is the nucleon interpolation field.  Note
that we take the quark source to be at the origin $\vec{x} = 0$,
$t = 0$ in the above expression. In the actual calculation, we take the source to be a smeared grid
source with a $Z_3$ noise. A diagram representing the three-point function is drawn in
\figref{fig:di}

The momentum fraction can then be obtained via the following ratio
\begin{equation}
  -\frac{1}{p} \frac{\trace [\Gamma_e C_3(t_1, t_2, \vec{p})]}{\trace [\Gamma_e
    C_2^N(t_2, \vec{p})]} \rightarrow \la x \ra, \quad t_2 \gg t_1 \gg
  0
\end{equation}
in which $\Gamma_e$ is the parity projection operator
$(1+\gamma_4)/2$.  To improve signal/noise ratio, we use the sum
method\cite{maiani_scalar_1987} by summing the insertion time $t_1$
over an appropriate range
\begin{equation}
  \label{eq:sum}
  -\frac{1}{p} \sum_{t_1 = 1}^{t_2-1} \frac{\trace [\Gamma_e C_3(t_1, t_2,
    \vec{p})]}{\trace [\Gamma_e C_2^N(t_2, \vec{p})]} \rightarrow \la
  x \ra t_2 + \mathrm{const.}
\end{equation}

We use the overlap fermion valence on $N_{f} = 2+1$  domain wall fermion configurations
by the RBC/UKQCD collaboration\cite{yamazaki_nucleon_2008}, for which the lattice
spacing is $a^{-1} = 1.77\,$GeV.  The light quark mass is $0.005$ in
lattice unit, which corresponds to $m_{\pi}=330\,$MeV.  For this work,
200 configurations are used.  In the valence sector, quark masses
$m_q = 0.00809$, $0.0135$, $0.016$, $0.0203$, and $0.0576$ are chosen
to be included.

On the choice of momentum for the two-point function, in principle,
one may use any finite momentum allowed on the lattice.  We choose
momenta $p = 2\pi/La$ and $4\pi/La$, and take an appropriate average to
acquire a final result.

\section{Lattice techniques}

\begin{figure}
  \centering
  \includegraphics[width=0.8\textwidth]{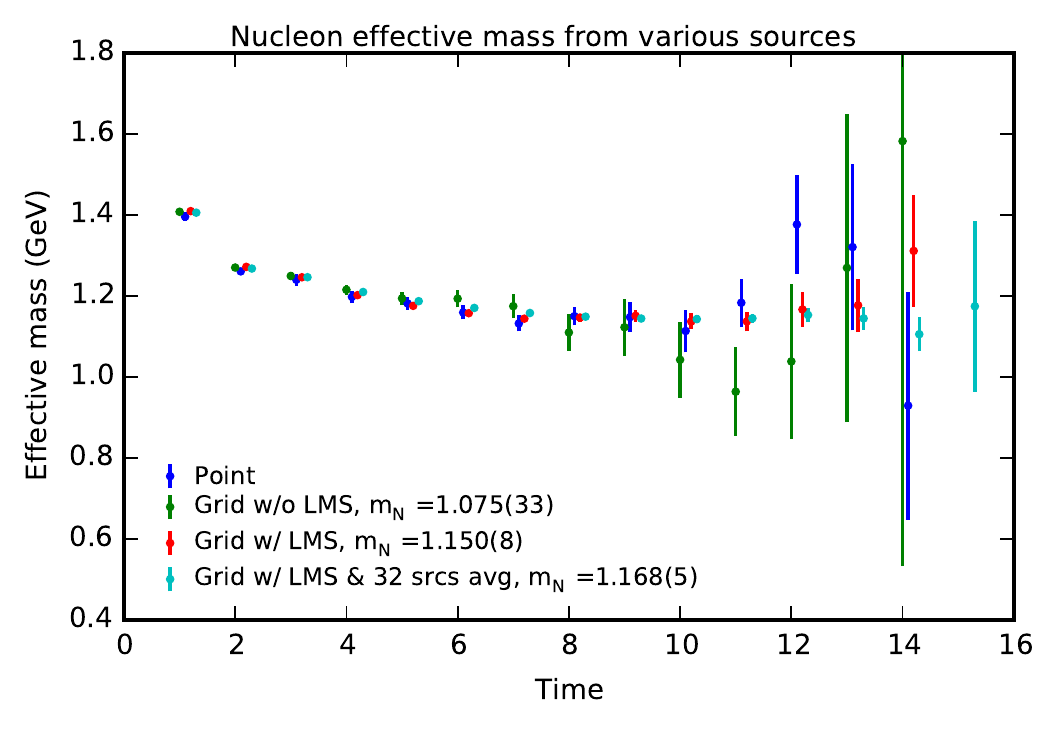}
  \caption{\label{fig:nucleon-compare} Comparison of nucleon effective
    mass at $m_{\pi} = 330\,$MeV plotted from different source schemes
    (as ordered in the legend: a single point, 8-grid without LMS,
    8-grid with LMS, and 8-grid with LMS from 32 sources per
    configuration as described in this paper).  In this graph, all
    sources are gaussian smeared. “Grid w/ LMS \& 32 srcs avg” is used
    in this work.}
\end{figure}

For the two-point correlation function, several techniques are applied
to improve the signal/noise ratio.  We use an 8-grid source with
gaussian smearing\cite{alexandrou_static_1994} to enhance the signal
of the ground state and $Z_3$ stochastic noise is used to tie the
three-quarks in each smeared source together to have a better overlap
with the nucleon.  Low mode
substitution\cite{li_overlap_2010,gong_strangeness_2013} is also
utilized; low mode contribution to the two-point function is
calculated exactly.  Furthermore, for each configuration, we use 32
sources, each of which is shifted in the time direction.  Since at
large $t$ where the nucleon effective mass plateau starts to appear,
high mode contribution is less important, the inversion of the quark
matrix with low-mode deflation is calculated with lower precision to
save inversion time.  \Figref{fig:nucleon-compare} shows the
difference in error bars of a point source, grid source, and grid
source with low mode substitution (all gaussian-smeared).  We should
mention that without low-mode substitution, the signal from the grid
source is even worse than that from the smeared point source, despite
having more statistics (also see \cite{li_overlap_2010}).

\begin{figure}
  \centering
  \includegraphics[width=0.7\textwidth]{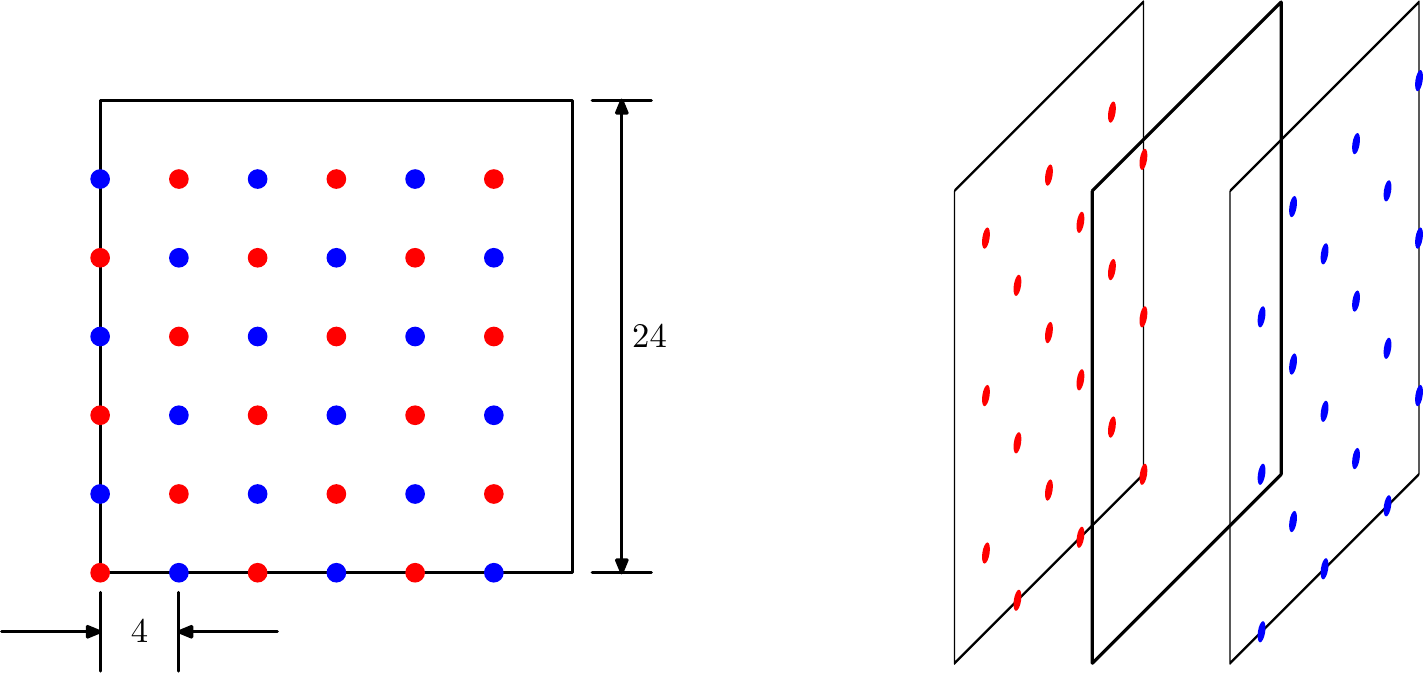}
  \caption{\label{fig:loop-source} Source scheme for the loop. The right figure
  shows the grid points on different time slices in each diluted inversion.}
\end{figure}

For the loop, we use a grid source with $Z_4$ noises in both spacial
and time direction, and the sources are arranged in a even-odd
fashion.  For each configuration, 32 sources are used.  Among all
these sources, the grid is shifted in spacial directions, so that
these sources together cover all time slices, and $1/4$ of the lattice
sites in any given time slice.  A sketch of this source scheme is
drawn in \figref{fig:loop-source}.  Similar to the two-point
functions, we utilize the low mode averaging (LMA)
technique\cite{foley_practical_2005}, and calculate the low mode
contribution to the loop exactly without any stochastic noise, and
high mode contribution is calculated with lower precision.

\section{Results}

\begin{figure}
  \centering
\begin{subfigure}{.5\textwidth}
  \centering
  \includegraphics[width=0.9\textwidth]{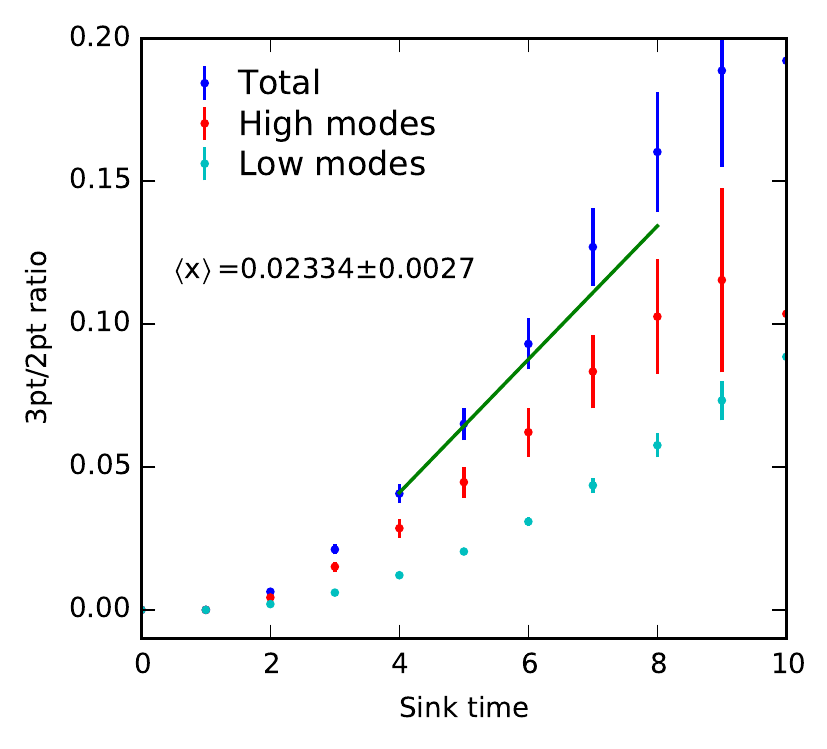}
  \caption{$\la x\ra$ slope in the u/d channel}
  \label{fig:slope-ud}
\end{subfigure}%
\begin{subfigure}{.5\textwidth}
  \centering
  \includegraphics[width=0.9\textwidth]{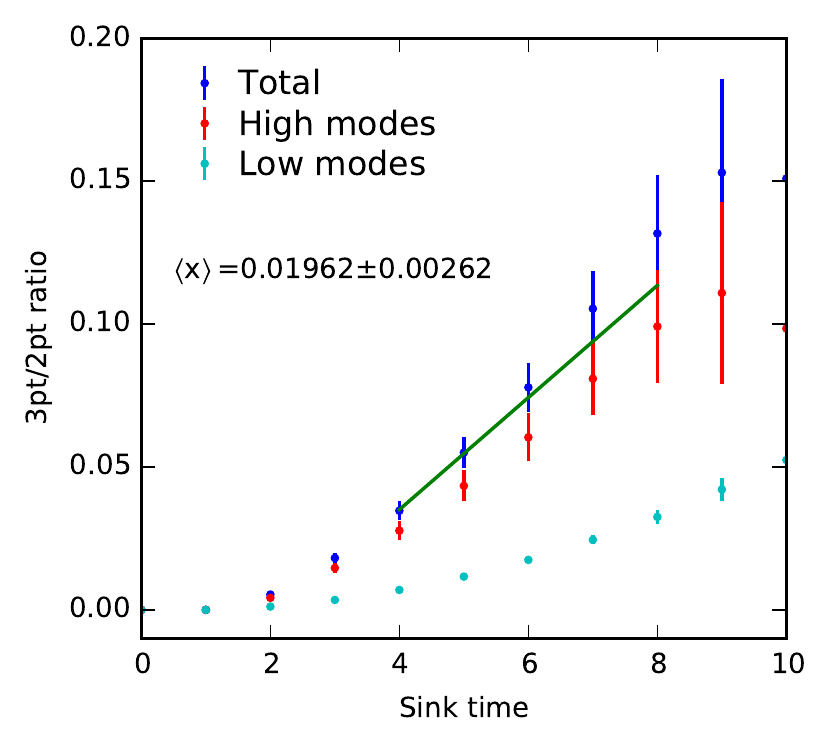}
  \caption{$\la x\ra$ slope in the strange channel}
  \label{fig:slope-s}
\end{subfigure}
\caption{3-point/2-point function ratio in the form of slope as a function of the
nucleon sink-source time separation from the sum method.}
\end{figure}

In order to improve the signal/noise ratio, we use the sum method to
calculate the physical quantities we want, i.e. $\la x\ra$, from the 3-point
function (\eqref{eq:sum}).  \Figref{fig:slope-ud} shows the signals in
the u/d channel, in which both the 2-point function and the loop are
with quark mass $0.016$ ($m_{\pi} = 330$ MeV); \figref{fig:slope-s} shows the
signal in the strange channel, in which the nucleon propagator is calculated with quark
mass $0.016$, and the loop with $0.0576$ which is at the strange quark mass.
All the techniques described above contribute to the resulting good signals. In the u/d
channel, we are able to reach a clear signal almost 9 sigmas away from zero.  Furthermore,
contribution from high modes and low modes are also plotted.  It is
evident that the high modes contribution is significant for
$\la x \ra$, hence we devote a large amount of computational time in order to
control its error.

\begin{figure}
  \centering
  \includegraphics[width=0.9\textwidth]{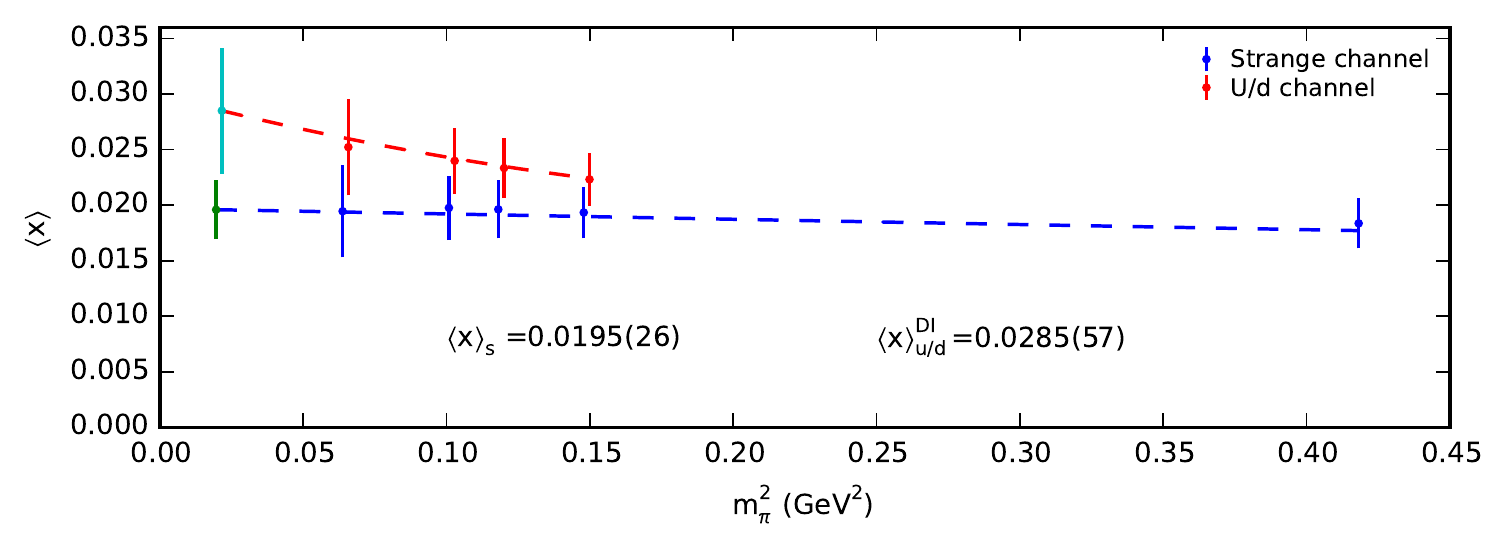}
  \caption{\label{fig:extra} $\la x \ra_{ud}$ and $\la x \ra_{s}$
    extrapolated to the physical value}
\end{figure}

We show in \figref{fig:extra} $\la x \ra_{u/d}^{\mathrm{DI}}$ and
$\la x \ra_{s}$ as a function of $m_{\pi}^2$ for several quark masses
in the nucleon propagator and the $u/d$ quark loop;  while the strange
quark mass in the loop is fixed. We then extrapolate to the physical
pion mass using the following functional forms
respectively 
\begin{equation}
  \la x \ra_{ud}^{\mathrm{DI}} = a_1 + a_2 m_{\pi}^2 + a_3 m_{\pi}^4, \quad \la
  x\ra_{s} = b_1 + b_2 m_{K}^2.
\end{equation}
in which the $a$'s and $b$'s are fitting parameters.  Thus we
obtain both $\la x \ra_{u/d}^{\mathrm{DI}}$ and $\la x \ra_{s}$ at the physical point
\begin{displaymath}
  \la x \ra_{u/d}^{\mathrm{DI}} = 0.0285(57), \quad  \la x \ra_{s} = 0.0195(26).
\end{displaymath}
They will be renormalized and matched to the $\overline{MS}$
scheme~\cite{glatzmaier_perturbative_2014}.

\begin{figure}
  \centering
  \includegraphics[width=0.8\textwidth]{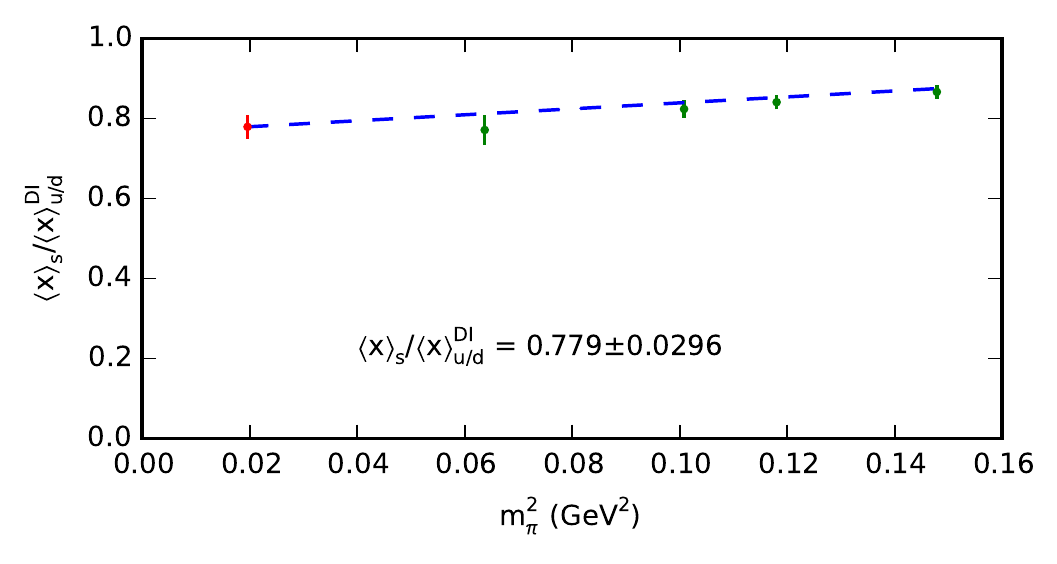}
  \caption{$\la x \ra_{s}/\la x \ra_{u/d}^{\mathrm{DI}}$ ratio
    extrapolated to the physical pion mass.}
  \label{fig:ratio}
\end{figure}

In \figref{fig:ratio}, we plot the ratio
$\la x \ra_{s} / \la x \ra_{u/d}^{\mathrm{DI}}$ for each quark mass, and take a
chiral extrapolation in terms of the linear $m_{\pi}^2$. The extrapolated
value is found to be
\begin{displaymath}
  \frac{\la x \ra_{s}}{\la x
    \ra_{u/d}^{\mathrm{DI}}} = 0.78(3).
\end{displaymath}

\section{Conclusion and Discussion}

\begin{figure}
  \centering
  \includegraphics[width=0.6\textwidth]{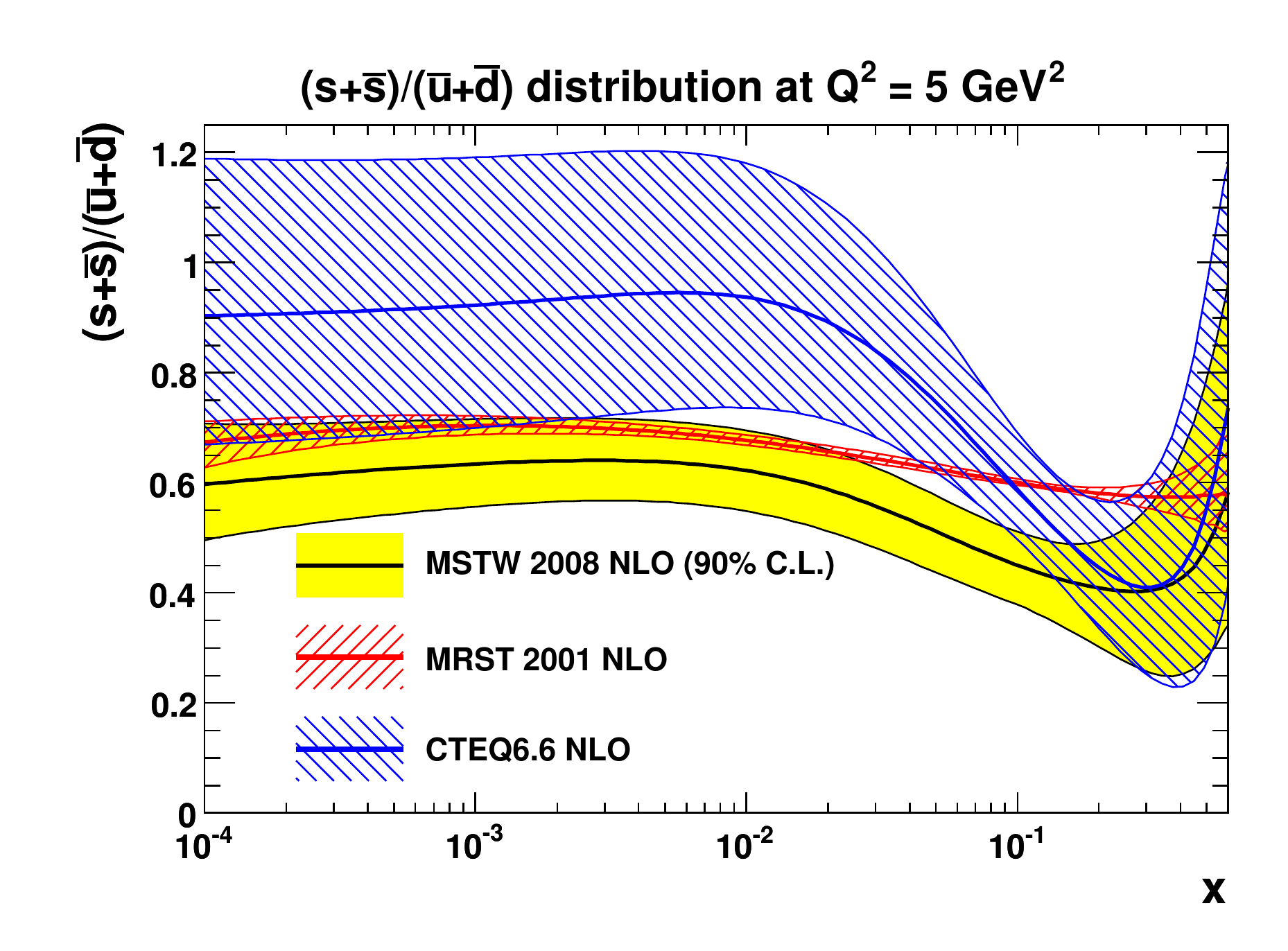}
  \caption{Ratio of $s + \bar{s}$ over $\bar{u} + \bar{d}$ versus $x$
    at $Q^2 = 5\,\mathrm{GeV}^2$ from various recent PDFs.}
  \label{fig:ratio-global}
\end{figure}

We have studied the momentum fraction carried by quarks in nucleon
with overlap fermions on $2+1$ flavors domain wall dynamical fermion
configurations.  We utilize an array of lattice techniques including
low mode substitution and low mode averaging, as well as grid source
with $Z_3$ and $Z_4$ noises to greatly improve the signal of both the
nucleon propagator and the quark loop.  We find clear signals for
$\la x \ra_{u/d}^{\mathrm{DI}}$ with a 5 sigma signal and
$\la x \ra_{s}$ with a 7 sigma signal at the physical pion mass.  It
is worth noting that our ratio result
$\la x \ra_{s}/\la x \ra_{u/d}^{\mathrm{DI}} = 0.78(3)$ is consistent
with the global analysis shown in
\figref{fig:ratio-global}\cite{martin_parton_2009}, given that the
``disconnected sea'' contribution dominates small $x$ region due to
its $x^{-1}$ dependence\cite{liu_connected-sea_2012,peng_flavor_2014},
and that the $(s + \bar{s}) / (\bar{u} + \bar{d})$ ratio is relatively
flat in small $x$ region.  We would also like to point out that our
result is in agreement with previous lattice calculations by T.~Doi et
al.~\cite{doi_strangeness_2008}\ and M.~Deka et
al.~\cite{m._deka_moments_2009}. This ratio can be used to better
constrain the global fitting of the parton distribution functions at
small $x$.

\bibliography{mylibrary}{}
\bibliographystyle{h-physrev}
\end{document}